\pdfoutput=1
\documentclass[superscriptaddress, 12pt, aps, prl,  preprint, longbibliography]{revtex4-1}
\usepackage{graphicx}
\usepackage{epstopdf}
\usepackage{float}
\usepackage{color}
\usepackage{amsmath, amssymb}

\begin{document}

\title{Nanoconfinement Facilitates Reactions of Carbon Dioxide in Supercritical Water}
\author{Nore Stolte}
\affiliation{Department of Physics, Hong Kong University of Science and Technology, Hong Kong, China}
\altaffiliation[Present address: ]{Lehrstuhl f\"ur Theoretische Chemie, Ruhr-Universit\"at Bochum, 44780 Bochum, Germany}
\author{Ding Pan}
\email{dingpan@ust.hk}
\affiliation{Department of Physics, Hong Kong University of Science and Technology, Hong Kong, China}
\affiliation{Department of Chemistry, Hong Kong University of Science and Technology, Hong Kong, China}
\affiliation{HKUST Fok Ying Tung Research Institute, Guangzhou, China}

\date{\today}
\begin{abstract}
The reactions of CO$_2$ in water under extreme pressure-temperature conditions are of great importance to the carbon storage and transport below Earth's surface, which substantially affect the carbon budget in the atmosphere.   
We applied ab initio molecular dynamics simulations to study aqueous carbon solutions nanoconfined by graphene and stishovite (SiO$_2$) at 10 GPa and 1000$\sim$1400 K. 
We found that CO$_2$(aq) reacts more in nanoconfinement than in bulk.
The stishovite-water interface makes the solutions more acidic, which shifts the chemical equilibria, and the interface chemistry also affects the reaction mechanisms.
Our findings suggest that CO$_2$(aq) in deep Earth may be more active than previously thought, and confining CO$_2$ and water in nanopores may enhance the efficiency of mineral carbonation.
\end{abstract}
\maketitle

\section{Introduction}
Aqueous fluids play a critical role in transporting carbon between Earth's surface and interior \cite{Peacock1990Fluid, Kelemen2015Reevaluating, manning2020subduction}, 
which is a substantial part of Earth's carbon cycle, with great implications for global climate and human's energy consumption.
It has long been assumed that aqueous carbon solutions under extreme pressure (P) and temperature (T) conditions are made by mixtures of neutral gas molecules \cite{Zhang2009Model}, e.g., H$_2$O, CO$_2$, CH$_4$; however, recent studies showed that important chemical reactions may occur between water and carbon species, resulting in significant amounts of ionic products, which may further participate in water-rock interactions and the formation of diamonds in Earth's interior \cite{Facq2014Insitu, Sverjensky2014Important, Pan2016Fate, Abramson2017Water-Carbon, Manning2018Fluids, Stolte2019Large}. 
Most of the previous studies focus on properties of aqueous carbon solutions in the bulk phase. 
In fact, aqueous solutions in deep Earth are often confined to the nanoscale in pores, grain boundaries, and fractures of Earth's materials \cite{gautam2017structure, Marquardt2018Structure, Huang2020Experimental}, where the physical and chemical properties of solutions may be dramatically different from those of bulk solutions.
Additionally, in carbon capture and sequestration efforts, CO$_2$ mineralization occurring in water trapped in porous rocks offers an efficient and secure method to permanently store carbon underground with low risk of return to the atmosphere \cite{Snaebjornsdottir2020Carbon}. 
The behavior of aqueous carbon solutions under nanoconfinement at extreme P-T conditions is of great importance to the deep carbon cycle and CO$_2$ storage, but is poorly understood on the molecular scale.

Previous studies reported that nanoconfinement substantially affects properties of water, e.g., equation of state \cite{Gunther2008Novel, Long2011Pressure, Hartkamp2012Study}, phase behavior \cite{Algara-Siller2015Square, Chen2016Two, Gao2018Phase}, dielectric constant \cite{Bonthuis2011Dielectric, Munoz-Santiburcio2017Nanoconfinement, Fumagalli2018Anomalously, Ruiz-Barragan2020Quantifying}, and diffusion \cite{Chiavazzo2014Scaling, Thompson2018Perspective, Phan2020Evidence}; as a result the reactivity of solutes under confinement may be very different from that in bulk solutions \cite{Munoz-Santiburcio2021Confinement}.
The dimensional reduction and increased fluid density may enhance reactions between small solutes in nanoconfinement \cite{Turner2001Effect, Santiso2005Adsorption}, whereas reactions involving large reactants or intermediates may be sterically hindered \cite{Munoz-Santiburcio2017Chemistry}.
Further,  the increase of the dielectric constant of nanoconfined water parallel to the confining surface leads to stabilization of charged aqueous reaction products \cite{Munoz-Santiburcio2017Chemistry}, causing the enhanced autodissociation of water \cite{Munoz-Santiburcio2017Nanoconfinement}. 
The solid-liquid interface also greatly affects 
the properties of confined aqueous solutions \cite{Phan2014Preferential}. Preferential adsorption of solutes at the confining interface may shift reaction equilibria. For example, in the production of methane from carbon dioxide at hydrothermal vent conditions (CO$_2$ + 4 H$_2$ $\rightleftharpoons$ CH$_4$ + 2 H$_2$O), hydrophilic pore surfaces adsorb water, favoring the production of methane \cite{Le2017Confinement}. 

Nanoconfinement and interface chemistry may both likely change the properties of aqueous carbon solutions, but a molecular understanding is lacking on how chemical speciation and reaction mechanisms are affected.
It was experimentally found that magnesite precipitates much faster in nanoscale water films than in bulk water \cite{miller2019anomalously}.
Because it is very challenging to study aqueous solutions under nanoconfinement in experiment, atomistic simulations are widely used. 
Many studies applied classical force fields \cite{Cole2010Supercritical, Phan2014Preferential, Chiavazzo2014Scaling, Phan2020Evidence}, which were usually designed for bulk solutions at ambient conditions; their accuracy at extreme conditions is not well tested.
As a comparison, ab initio molecular dynamics (AIMD) simulations do not rely on experimental input or empirical parameters \cite{car1985unified, galli1991ab, marx2009ab}. We solve the many-body electronic structure numerically, so the breaking and forming of chemical bonds, electronic polarizability, and charge transfer are all treated at the quantum mechanical level \cite{gygi2005ab, marx2009ab}. The AIMD method is widely considered as one of the most reliable methods to make predictions, and many simulations results were later confirmed by experiments \cite{gygi2005ab, marx2009ab}.

Here, we performed extensively long AIMD simulations to study CO$_2$(aq) solutions nanoconfined by graphene and stishovite (SiO$_2$) at 10 GPa and 1000$\sim$1400 K. 
These P-T conditions are typically found in Earth's upper mantle.
We compared the CO$_2$(aq) reactions in nanoconfinement with those in the bulk solutions,
and examined how weak and strong interactions between confining walls and confined solutions may affect chemical speciation and reaction mechanisms.  
Our work is of great importance to the carbon transformation in deep Earth, and also helps us to understand atomistic mechanisms of CO$_2$ mineralization in the carbon capture and storage. 

\section{Results and discussion}
\begin{figure}[H]
\centering
\includegraphics[width=0.5\textwidth]{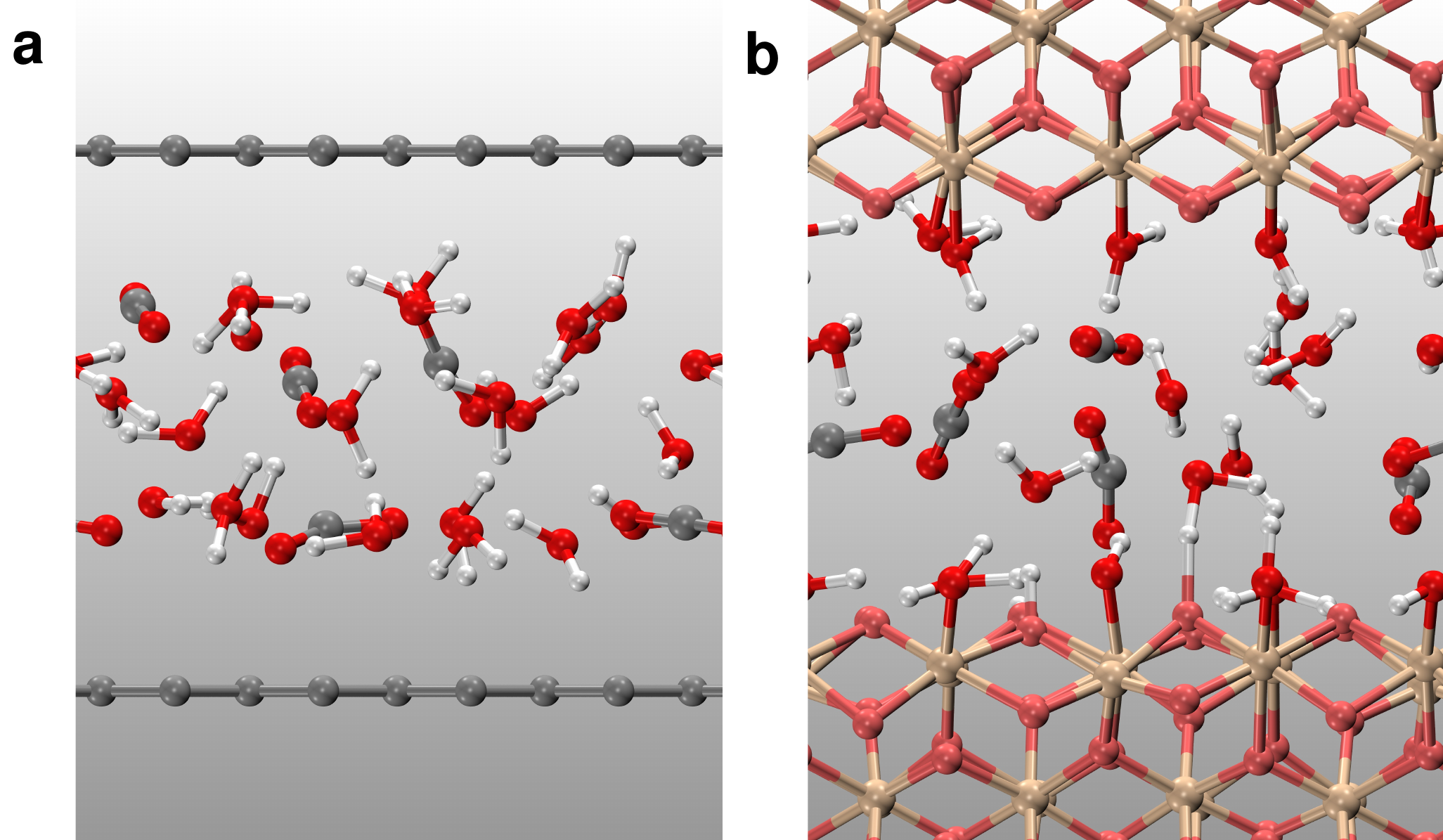}
\caption{Snapshots of ab initio molecular dynamics simulations under confinement. (\textbf{a}) CO$_2$(aq) confined by graphene sheets. 
(\textbf{b}) CO$_2$ confined by cleaved stishovite (SiO$_2$) (100) slabs. 
Red balls represent oxygen atoms in solutions, and pink balls are oxygen atoms in SiO$_2$. 
Gray, white, and yellow balls are carbon, hydrogen, and silicon atoms, respectively.}
\label{setup}
\end{figure}

We first studied CO$_2$(aq) solutions confined by two graphene sheets
at $\sim$10 GPa, and 1000$\sim$1400 K (see Fig. \ref{setup}(a)).
The graphene sheet separation was 9.0 and 9.2 {\AA} at 1000 K and 1400 K, respectively.
The graphene sheets were modeled using a distance-dependent potential acting on the carbon and oxygen atoms, which was fitted to the interaction energies calculated using diffusion quantum Monte Carlo \cite{Brandenburg2019Physisorption} and van der Waals density functional theory \cite{Takeuchi2017Adsorption} (see Fig. S1 in the supporting information).
We calculated the pressure of the confined solutions  parallel to the graphene sheets, which are $\sim$10 GPa, consistent with the equation of state (see Table SI in the supporting information).
In addition, We also used the atom number density profiles to calculate the actual volumes that aqueous carbon solutions occupy, and then applied the equation of state of CO$_2$ and water mixtures to obtain the pressures \cite{Duan2006Equation}. 

We directly dissolved the CO$_2$ molecules in the supercritical water, and the initial mole fraction of CO$_2$(aq) is 0.185.
The CO$_2$ molecules reacted frequently with water, and
we performed long AIMD simulations until the concentrations of carbon species reached equilibrium, which took  180$\sim$480 ps (see Fig. S2 in the supporting information).
Initially, the reaction between CO$_2$(aq) and H$_2$O produces bicarbonate ions (HCO$_3^-$):  
\begin{equation} \label{reaction-co2} \mathrm{ CO_2(aq) + 2 H_2O \rightleftharpoons HCO_3^- + H_3O^+ } . 
\end{equation} 
This reaction may occur in one step, or involves dissociation of water so that OH$^-$ can react with CO$_2$(aq) to form HCO$_3^-$(aq).
The generated bicarbonate ion may further accept a proton to become a carbonic acid molecule (H$_2$CO$_3$(aq)), or may lose a proton to become a carbonate ion (CO$_3^{2-}$(aq)). The major carbon  species in the solutions are CO$_2$, CO$_3^{2-}$, HCO$_3^-$, and H$_2$CO$_3$.

\begin{figure}[H]
\centering
\includegraphics[width=0.5\textwidth]{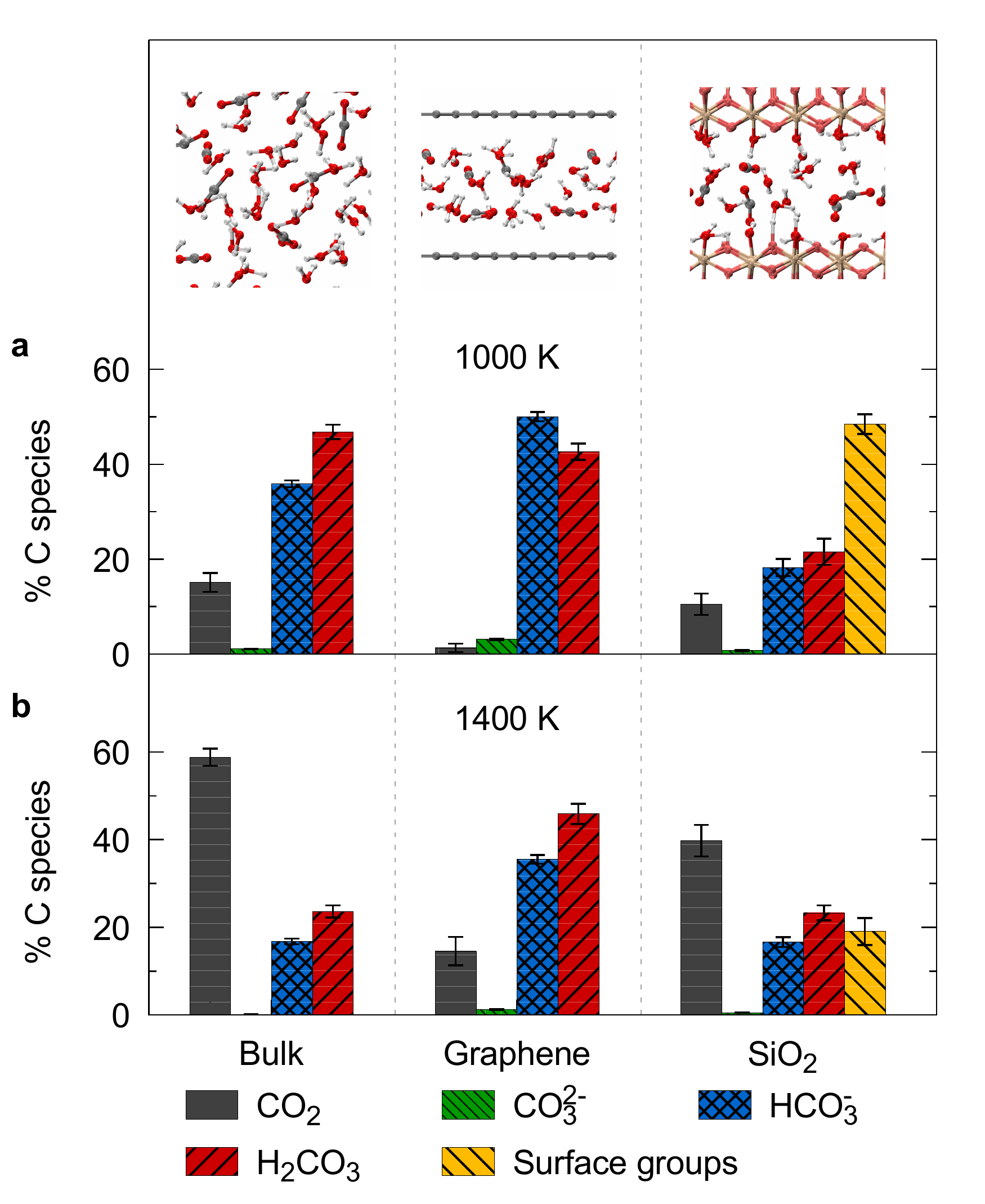}
\caption{Mole percents of carbon species in the CO$_2$(aq) solutions in the bulk and nanoconfined by graphene and stishovite (SiO$_2$) at chemical equilibrium. 
The initial mole fraction of CO$_2$(aq) is 0.185. The pressures in all solutions are $\sim$10 GPa. The temperatures are (\textbf{a}) 1000 K and (\textbf{b}) 1400 K. 
The data of bulk solutions in (\textbf{a}) are from Ref. \cite{Stolte2019Large}, and the bulk data in (\textbf{b}) were interpolated using the simulation results in Ref. \cite{Stolte2019Large}.
Uncertainties were obtained using the blocking method \cite{Flyvbjerg1989Error}.}
\label{overall}
\end{figure}

We compared the chemical speciation of the solutions under nanoconfinement and in the bulk phase at the same P-T conditions \cite{Stolte2019Large}. Fig. \ref{overall} shows that 
at 1000 K, the mole percent of CO$_2$(aq) in total dissolved carbon under nanoconfinement is 1.3$\pm$0.9\%,
whereas it is 15.2$\pm$2.0\% in the bulk solution.
The mole percent of HCO$_3^-$ under the nanoconfinement (50.0$\pm$1.0\%) is higher than that in the bulk solution (35.9$\pm$0.7\%), and the concentrations of H$_2$CO$_3$(aq) are similar (42.7$\pm$1.7\% vs. 46.8$\pm$1.5\%). 
With increasing temperature from 1000 K to 1400 K, the mole percents of CO$_2$(aq) under nanoconfinement and in the bulk solution increase to
14.5$\pm$3.2\% and 58.8$\pm$2.0\%, respectively.
The equilibrium concentrations of CO$_2$(aq) in the nanoconfined solutions are lower than those in the bulk solutions at the two temperature conditions studied here, suggesting that
nanoconfinement promotes the 
CO$_2$(aq) reactions.
When increasing temperature along the isobar, 
due to thermal entropy effects, small molecules like CO$_2$(aq) are more favored.
We did not see obvious difference in reaction rates between the nanoconfined and bulk solutions.

\begin{figure}[H]
\centering
\includegraphics[width=0.6\textwidth]{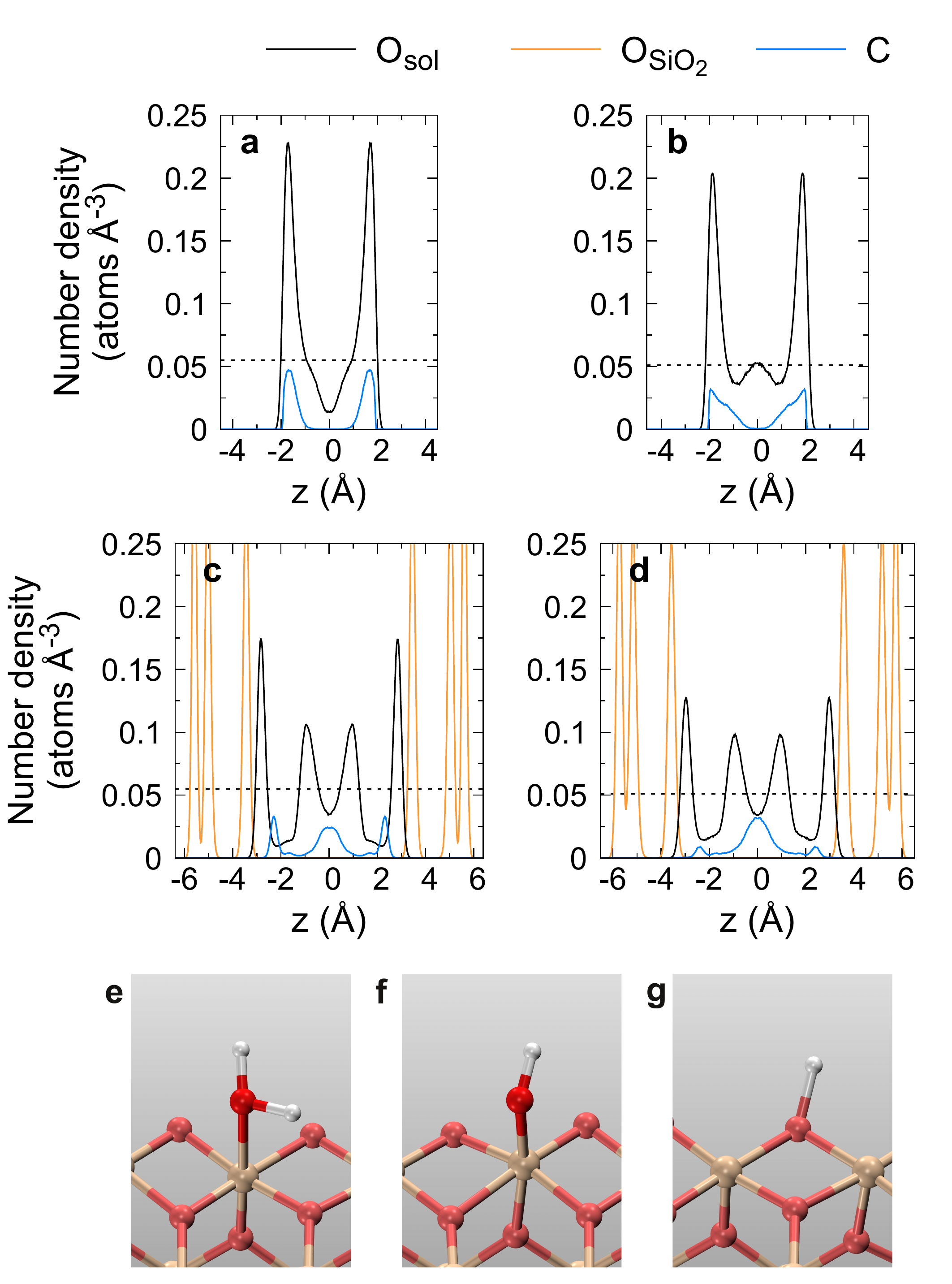}
\caption{Number density profiles of oxygen atoms (O$_{\mathrm{sol}}$) and carbon atoms along the $z$ axis, normal to the confining surfaces. 
 (\textbf{a}) and (\textbf{b}) show the solutions under graphene confinement at 1000 and 1400 K, respectively.
(\textbf{c}) and (\textbf{d}) show
the solutions under stishovite confinement at 1000 and 1400 K, respectively.
O$_{\mathrm{SiO_2}}$ refers to the oxygen atoms in stishovite. 
The pressures in all solutions are $\sim$10 GPa.
The initial mole fraction of CO$_2$(aq) is 0.185.
The center of confined fluids is set at $z = 0$, and the density distributions have been symmetrized.
The horizontal black dashed lines represent the oxygen density in bulk solutions at the corresponding P-T conditions \cite{Duan2006Equation}.
 (\textbf{e}), (\textbf{f}), and  (\textbf{g}) show the H$_2$O molecule, the OH$^-$ and H$^+$ ions bonded to the stishovite (100) surface, respectively. }
\label{density}
\end{figure}

\begin{figure}[H]
\centering
\includegraphics[width=0.6\textwidth]{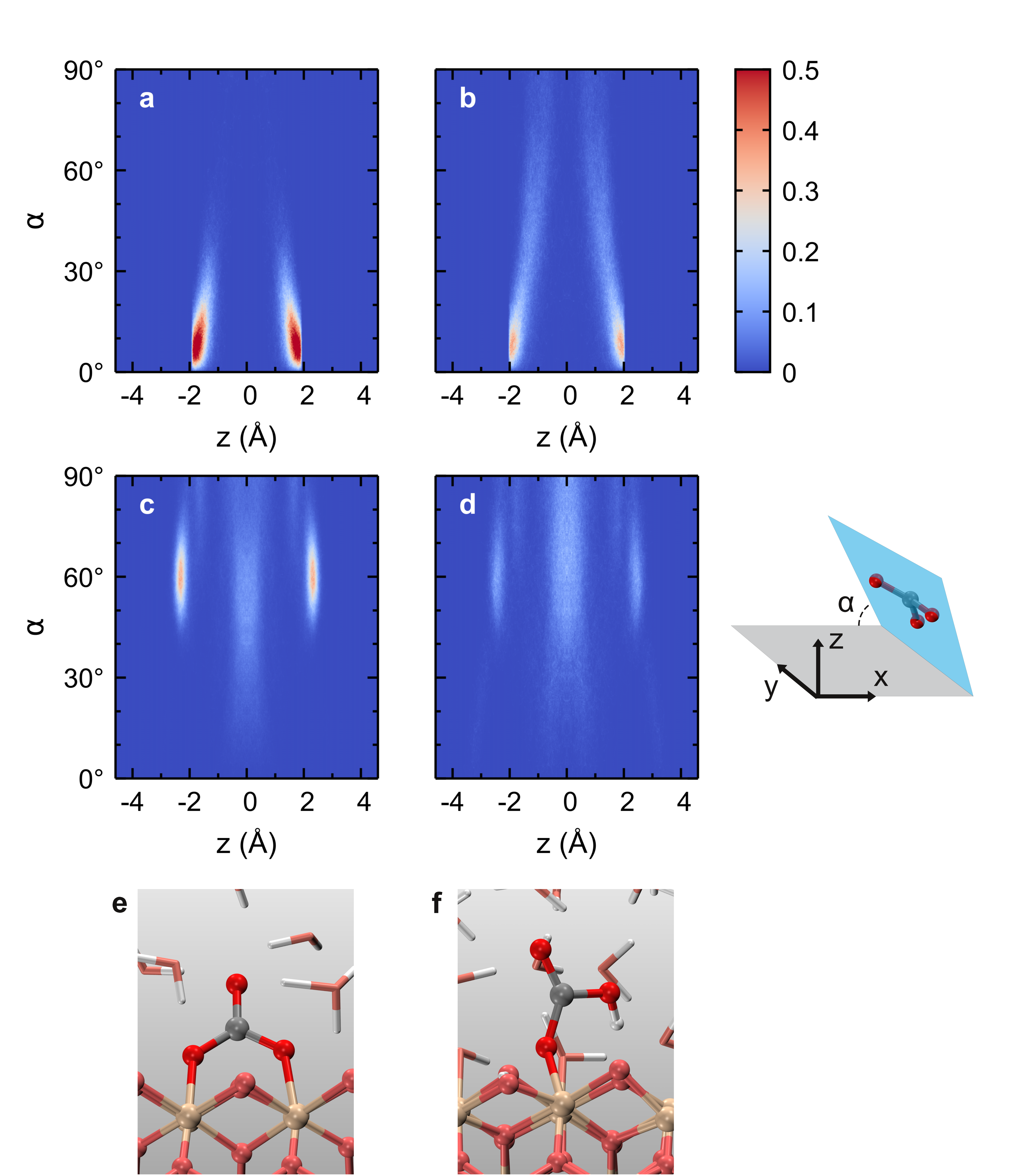}
\caption{Orientation distribution of sp$^2$ carbon species in CO$_2$(aq) solutions.
The dihedral angle $\alpha$ is between the confinement interface and the plane defined by the three oxygen atoms in sp$^2$ carbon species.
(\textbf{a}) and (\textbf{b}) show the solutions under graphene confinement at 1000 and 1400 K, respectively.
(\textbf{c}) and (\textbf{d}) show
the solutions under stishovite confinement at 1000 and 1400 K, respectively.
The pressure in all solutions are $\sim$10 GPa. 
The initial mole fraction of CO$_2$(aq) is 0.185.
The center of confined fluids is set at $z= 0$, and the angle distributions have been symmetrized.
(\textbf{e}) and (\textbf{f})
show the CO$_3^{2-}$ and HCO$_3^-$ ions adsorbed on the stishovite (100) surface, respectively.
}
\label{co3-config}
\end{figure}

To understand why the nanoconfinement promotes the CO$_2$(aq) reactions, 
we analyzed the water structure in Fig. \ref{density}.
In the graphene-confined solutions, there are two sharp density peaks for oxygen atoms, corresponding to two water layers (Fig. \ref{density}(a) and (b)).
We found that the carbon-containing ions and molecules in these two layers tend to align parallel to the graphene sheets (see Fig. \ref{co3-config}), and CO$_2$(aq) mostly reacts with water molecules in the same layer (Fig. S4 in the supporting information).
Nanoconfinement increases the probability of reactive encounters between CO$_2$(aq) and solvent molecules, as diffusion is restricted to two dimensions  \cite{Munoz-Santiburcio2017Chemistry}.
It has been reported that the dielectric constant of nanoconfined water in the direction parallel to the confining surfaces, $\epsilon_{\parallel}$, increases significantly compared to the bulk value, and consistently 
water molecules dissociate more easily under nanoconfinement \cite{Munoz-Santiburcio2017Nanoconfinement, Ruiz-Barragan2020Quantifying}.
Here, the produced OH$^-$ ions from the water self-ionization may readily react with CO$_2$(aq) (reaction (\ref{reaction-co2})), and
the enhancement of $\epsilon_{\parallel}$ also further stabilizes the generated HCO$_3^-$ and CO$_3^{2-}$ ions.
As a result, more CO$_2$(aq) molecules react under nanoconfinement than in the bulk.

After studying the effects of graphene nanoconfinement, we turned to the confinement by a realistic mineral in deep Earth, stishovite, which is a stable phase of SiO$_2$ (space group: P4$_2$/mnm) at the P-T conditions studied here \cite{akaogi1995thermodynamic, zhang1996situ} and a major component of subducted oceanic crust \cite{aoki2004density}, playing a substantial role in transporting water into Earth's mantle \cite{Lin2020Evidence}.
We exposed the cleaved stishovite (100) face, one of the low-energy surfaces \cite{Feya2018Tetrahedral}, to the carbon solutions as shown in Fig. \ref{setup}(b).
We carried out constant-pressure (NPT) simulations to keep the pressure perpendicular to the solid-liquid interface at $\sim$10 GPa, and then we found that the distance between the outermost oxygen atoms in two stishovite (100) surfaces is $\sim$7 {\AA} (see Table S1 in the supporting information).

Fig. \ref{overall} shows the chemical speciation of aqueous carbon solutions under the stishovite confinement at $\sim$10 GPa and 1000$\sim$1400 K.
We found that at chemical equilibrium, 48.5$\pm$2.1\% (1000 K) and 19.1$\pm$3.1\% (1400 K) of carbon species, mostly HCO$_3^-$ and CO$_3^{-2}$, are bonded to the stishovite surfaces, unlike in the graphene-confined solutions. In the atomic density profiles shown in Fig. \ref{density}(c) and (d), there are oxygen and carbon density peaks near the SiO$_2$ surfaces, where the oxygen and carbon atoms come from the solutions, indicating that the solid-liquid interface plays an important role.

\begin{figure}[H]
\centering
\includegraphics[width=0.5\textwidth]{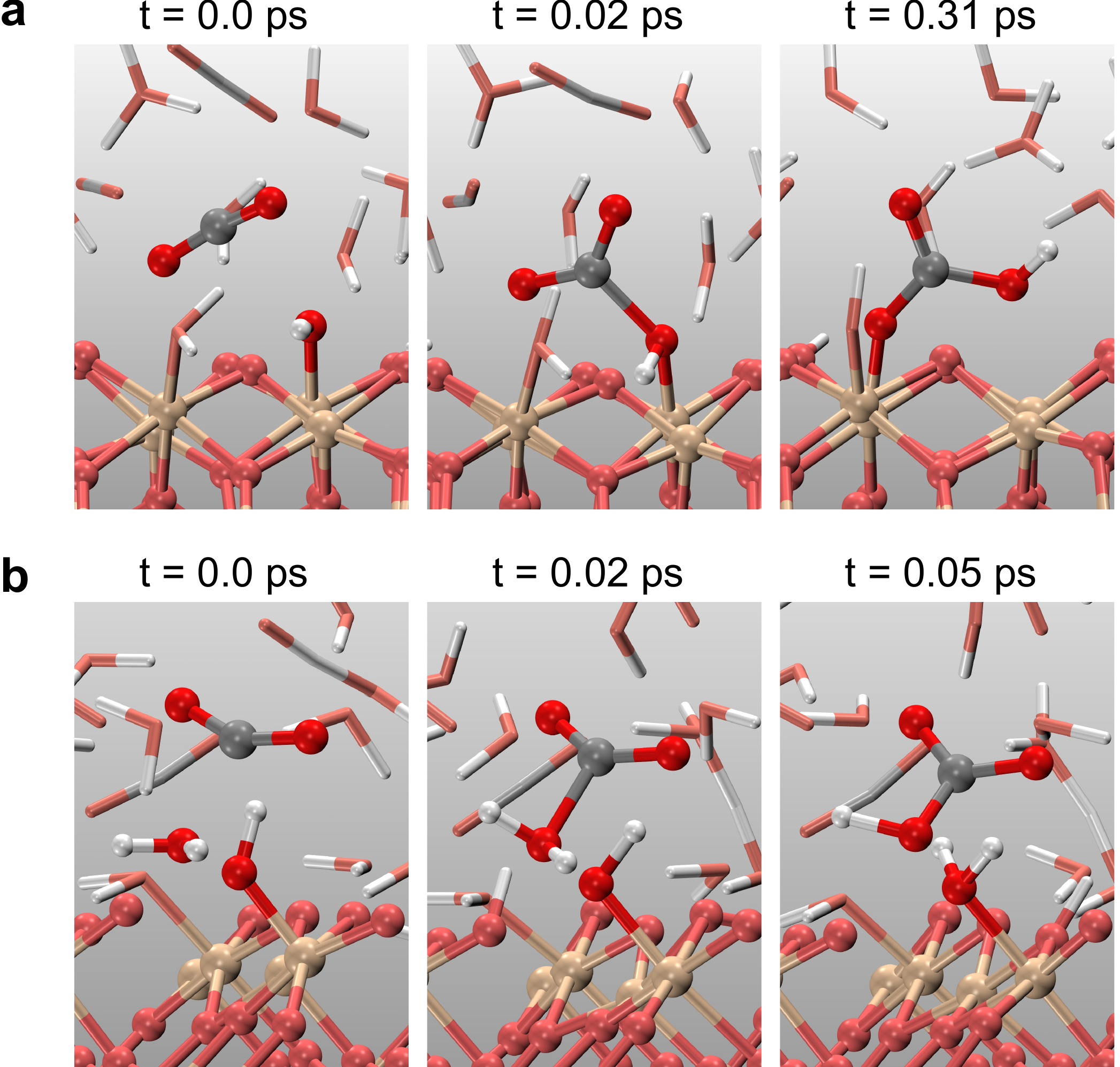}
\caption{Reaction of CO$_2$(aq) catalyzed by the stishovite (100) surface. (\textbf{a}) shows the formation of HCO$_3^-$ at the interface.
(\textbf{b}) shows the proton released from the reaction between CO$_2$(aq) and water is accepted by the silanol (Si-OH) surface group.
}
\label{reactions-stishovite}
\end{figure}

In bulk stishovite crystals, silicon atoms are octahedrally coordinated, and oxygen atoms are trigonally coordinated, whereas at the cleaved stishovite (100) surface, silicon atoms form bonds with five oxygen atoms, and each oxygen atom bridges two undercoordinated silicon atoms.
Water molecules may dissociate or bond at the stishovite surface.
The hydroxide ion (OH$^-$) from water dissociation can bond to an undercoordinated silicon atom to form a silanol (Si-OH) group, and the extra proton can bond with the surface oxygen atom to become a Si-(OH$^+$)-Si bridge (see Fig. \ref{density}({e}--{g})). Similar hydroxylation occurs at the quartz (1000) surface \cite{Adeagbo2008Transport, Ledyastuti2013First-principles}.
The hydroxyl groups at the stishovite surface may react with CO$_2$(aq) to form HCO$_3^-$ (Fig. \ref{reactions-stishovite}(a)). 
The surface hydroxyl groups or the undercoordinated oxygen atoms may also accept protons released in reaction (\ref{reaction-co2}), driving the reaction forward (Fig. \ref{reactions-stishovite}(b)).

We analyzed the spatial orientation of the sp$^2$ carbon species such as CO$_3^{2-}$(aq), HCO$_3^-$(aq), and H$_2$CO$_3$(aq) in the stishovite-confined solutions in Fig. \ref{co3-config}(c) and (d). We found that the molecular plane of the sp$^2$ carbon species
bonded to the stishovite surface 
tends to form an angle of $\sim$60$^{\circ}$ or $\sim$90$^{\circ}$ with the solid-liquid interface plane, dramatically different from the orientation of carbon species in the graphene-confined solutions.
When the angle is $\sim$60$^{\circ}$, 
the carbon species has two Si-O bonds by straddling two silicone atoms (Fig. \ref{co3-config}(e)), while with the angle of $\sim$90$^{\circ}$ the carbon species forms only one Si-O bond (Fig. \ref{co3-config}(f)).
The strong solid-liquid interaction substantially affects the molecular structure of the confined carbon solutions. 

After analyzing the carbon species at the solid-liquid interface, we investigated 
the carbon species not bonded to the stishovite surface, i.e., fully dissolved in the stishovite-confined solutions.
Fig. \ref{overall} show that at 1000 K, the mole percent of dissolved CO$_2$(aq) is 10.5$\pm$2.3\%, which is larger than 1.3$\pm$0.9\% in the graphene-confined solution, and slightly smaller than 15.1$\pm$2.0\% in the bulk solution.
At 1400 K, the mole percent of dissolved CO$_2$(aq) in the stishovite-confined solution (39.8$\pm$3.6\%) is also between those in the graphene-confined (14.5$\pm$3.2\%) and bulk (58.8$\pm$2.0\%) solutions.

Both hydroxide ions and protons may be chemically adsorbed on the SiO$_2$ surface, 
which affects the acidity of carbon solutions.
Considering that the pH value of neutral water is no longer 7 under extreme P-T conditions,
we calculated the difference between pH and pOH to quantify the acidity of solutions \cite{Dettori2020Carbon}:
\begin{equation}
f = \mathrm{pH - pOH} = - \log_{10} \left( \dfrac{\mathrm{[H_3O^+]}}{\mathrm{[OH^-]}}\right),
\end{equation}
where $\mathrm{[H_3O^+]}$ and $\mathrm{[OH^-]}$ are the concentrations of hydronium and hydroxide ions, respectively.
Because CO$_2$(aq) reacts with water to generate H$_3$O$^+$, 
the solutions studied here are all acidic, i.e., $f<0$, as shown in Table \ref{acidity}.
The interesting finding is that the $f$ value of stishovite-confined solutions is more negative than that of graphene-confined solutions at the same P-T conditions, which means that
the former is more acidic than the latter, even though less CO$_2$(aq) reacts in the stishovite-confined solutions.
We have discussed that the stishovite surface adsorbs the hydroxide ions and protons from solutions. Additionally, our AIMD trajectories show that
the SiO$_2$ surface favors the adsorption of OH$^-$ over that of H$^+$
(Fig. S5 in the supporting information); 
as a result, the stishovite-confined solutions are more acidic than the graphene-confined ones at the same P-T conditions.
Increasing the concentration of H$_3$O$^+$ shifts the equilibrium of reaction (\ref{reaction-co2}) towards the left, so there is more CO$_2$(aq) in the solutions. 

The nanoconfinement enhances $\epsilon_{\parallel}$, which stabilized charged ions, so in both graphene- and stishovite-confined solutions, more CO$_2$(aq) reacts.
However, it has been reported that $\epsilon_{\parallel}$ near the hydrophobic surface increases more than near the hydrophilic surface, because the motion of water molecules are more hindered at the the hydrophilic surface \cite{Bonthuis2011Dielectric}.
Considering that the stishovite surface is more hydrophilic than graphene, charged ions are less stabilized, so we found more CO$_2$(aq) in the stishovite-confined solutions than in the graphene-confined solutions. 
Therefore, the CO$_2$ concentration increase in the stishovite-confined solutions is a combined result of the hydrophilic confinement and the adsorption preference of OH$^-$ on the stishovite (100) surface.

In our simulations, we used the semilocal Perdew-Burke-Ernzerhof (PBE) exchange-correlation (xc) functional \cite{Perdew1996Generalized}, which was reported insufficient to describe aqueous systems at ambient conditions \cite{gillan2016perspective};  
however, our previous studies showed that PBE performed better for the equation of state  and dielectric properties of water \cite{Pan2013Dielectric,Pan2014Refractive} and the carbon speciation in water \cite{Pan2016Fate} at extreme P-T conditions than at ambient conditions.
Particularly, we compared the simulations using PBE and a hybrid xc functional, PBE0 \cite{adamo1999toward}.
For an aqueous carbon solution at $\sim$11 GPa and 1000 K, whose initial mole fraction of CO$_2$(aq) is 0.016, both PBE and PBE0 suggest that HCO$_3^-$ is the dominant carbon species, and its mole percents are 79.8\% and 75.0\%, respectively \cite{Pan2016Fate}.

\section{Conclusion}

We performed extensively long AIMD simulations to study 
the chemical reactions and speciation of aqueous carbon solutions nanoconfined by graphene and stishovite at 10 GPa and 1000$\sim$1400 K. We found that the graphene nanoconfinement promotes the CO$_2$(aq) reactions. 
When graphene is replaced by stishovite, 
less CO$_2$(aq) reacts, but still more than in the bulk solutions.
We found that contacting the stishovite (100) surface makes the solutions more acidic, which shifts the chemical equilibria, although 
the stishovite surface also catalyzes the CO$_2$(aq) reactions by adsorbing HCO$_3^-$ and H$^+$.

The enhanced reactivity of CO$_2$(aq) in nanoconfinement has important implications for carbon transport and fluid-rock interactions in deep Earth.
Aqueous fluids located at grain boundaries in minerals can either exist in isolated fluid-filled pores, or form a connected network of channels along grains facilitating fluid transport \cite{Marquardt2018Structure}.
It is known that adding molecular CO$_2$(aq) to water may increase the rock-fluid-rock dihedral angle $\theta$, which inhibits fluid flow \cite{Watson1987Fluids, Holness1992Equilibrium}.
However, our study shows that CO$_2$(aq) reacts with water under nanoconfinement and could react with the solid interface, which may decrease $\theta$ and promote interconnectivity of fluids \cite{Huang2020Experimental}.
Our study also sheds light on atomistic mechanisms of CO$_2$ storage through mineral carbonation. 
CO$_2$ reacts more in nanoconfined water, which benefits the CO$_2$ mineralization. 
If we choose minerals with larger points of zero charge than that of SiO$_2$, such as forsterite \cite{Wogelius1991Olivine} and magnesium oxide \cite{Kosmulski2016Isoelectric}, the CO$_2$ reactivity may be further enhanced.

\section*{Methods}
We carried out Born-Oppenheimer ab initio molecular dynamics using the Qbox package \cite{Gygi2008Architecture}. We used periodic boundary conditions and employed plane-wave basis sets and norm-conserving pseudopotentials \cite{Hamann1979Norm-conserving, Vanderbilt1985Optimally}, with a plane-wave cutoff of 85 Ry. The cutoff was increased to 145 Ry for pressure calculations. We applied density functional theory and the PBE exchange-correlation functional \cite{Perdew1996Generalized}. We sampled the Brillouin zone at the $\Gamma$ point. We performed AIMD simulations in the canonical, i.e., NVT, ensemble. Stochastic velocity rescaling was used to control the temperature \cite{Bussi2007Canonical}, with a damping factor of 24.2 fs. We replaced hydrogen by deuterium to use a large time step of 0.24 fs in the simulations, but still referred to these atoms as H atoms.

We ran simulations for 180--480 ps after 20 ps equilibration to reach chemical equilibrium. We analyzed the AIMD trajectories to determine the nature of carbon-containing molecules.
For each carbon atom, we searched for the three nearest oxygen atoms, and sorted the C-O distances in increasing order.
If the difference between the second and third C-O distance is less than 0.4 {\AA}, the carbon species is a CO$_3^{2-}$ ion; otherwise, it is CO$_2$.
Hydrogen atoms were considered being bonded to their nearest-neighbor oxygen atoms.
For the solutions confined by stishovite, the oxygen atoms were considered bonded to silicon atoms when the interatomic distance fell within the first peak of the Si-O$_{\mathrm{aq}}$ radial distribution function (RDF), i.e., 2.6 {\AA} (Fig. S3 in the supporting information).

\section*{Acknowledgements}
N.S. acknowledges the Hong Kong Ph.D. Fellowship Scheme. 
D.P. acknowledges support from the Croucher Foundation through the Croucher Innovation Award, Hong Kong Research Grants Council (Projects ECS-26305017, GRF-16307618, and GRF-16306621), National Natural Science Foundation of China (Project 11774072 and Excellent Young Scientists Fund), and the Alfred P. Sloan Foundation through the Deep Carbon Observatory.
Part of this work was carried out using computational resources from the National Supercomputer Center in Guangzhou, China.

\section*{Contributions}
D.P. designed the research. Calculations were performed by N.S. All authors contributed to the analysis and discussion of the data and the writing of the manuscript.

\bibliography{ref}
\clearpage

\begin{table}[H]
\caption{The acidity $f$ of carbon aqueous solutions: $f=\mathrm{pH-pOH}$.
The initial mole fraction of CO$_2$(aq) is 0.185, and the pressures in the solutions are about 10 GPa. Uncertainties are obtained using the blocking method \cite{Flyvbjerg1989Error}.}
\label{acidity}
\centerline{
\begin{tabular}{ l  l  c  }
\hline 
\hline
T $\qquad$ $\qquad$ & Confinement $\qquad$     & $\qquad$ $f$ $\qquad$    \\
\hline
1000 K & Graphene                & -1.24 $\pm$ 0.02 \\
\cline{2-3}
                           & Stishovite              & -1.46 $\pm$ 0.02 \\
\hline
1400 K & Graphene                & -0.94 $\pm$ 0.05 \\
\cline{2-3}
                           & Stishovite              & -1.13 $\pm$ 0.03 \\
\hline
\hline
\end{tabular}
}
\end{table}

\end{document}


\title{Supporting Information for:\\ Nanoconfinement Facilitates Reactions of Carbon Dioxide in Supercritical Water}

\author{Nore Stolte}
\affiliation{Department of Physics, Hong Kong University of Science and Technology, Hong Kong, China}
\altaffiliation[Present address: ]{Lehrstuhl f\"ur Theoretische Chemie, Ruhr-Universit\"at Bochum, 44780 Bochum, Germany}
\author{Ding Pan}
\email{dingpan@ust.hk}
\affiliation{Department of Physics, Hong Kong University of Science and Technology, Hong Kong, China}
\affiliation{Department of Chemistry, Hong Kong University of Science and Technology, Hong Kong, China}
\affiliation{HKUST Fok Ying Tung Research Institute, Guangzhou, China}

\maketitle

\begin{figure}[ht]
\centering
\includegraphics[width=0.5\textwidth]{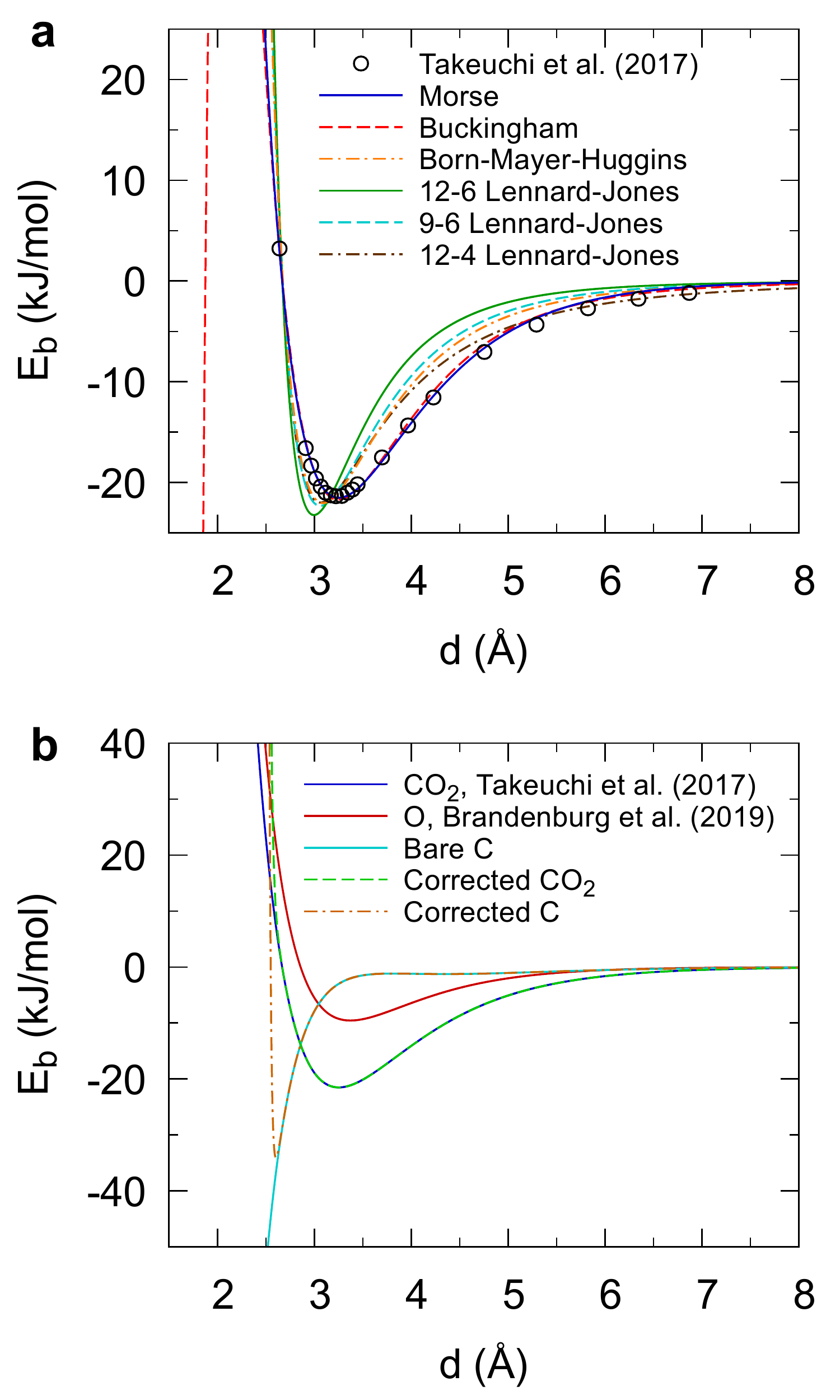}
\caption{Graphene-carbon and graphene-oxygen interactions in molecular dynamics simulations. $d$ is the distance between graphene and atoms. (\textbf{a}) Several force potentials are fitted to the data from the study on the graphene-CO$_2$ interaction by Takeuchi et al. \cite{Takeuchi2017Adsorption}.
The Morse potential is given by Eq. \ref{g-CO2}.
(\textbf{b}) 
The graphene-oxygen interaction (i.e., graphene-H$_2$O interaction) is from Brandenburg et al.'s study \cite{Brandenburg2019Physisorption}. The resulting bare graphene-carbon (Bare C) interaction is given by Eq. \ref{bare_C}. 
The graphene-carbon and graphene-CO$_2$ interactions after adding the repulsive term (Eq. \ref{repulsive}) are shown by the red dot-dashed (Corrected C) and blue dashed (Corrected CO$_2$) lines, respectively.
}
\label{potential}
\end{figure}

We modeled graphene in contact with solutions by defining a force potential that acts on carbon and oxygen atoms in solutions as a function of the perpendicular distance $d$ between 
graphene and atoms.
The graphene-oxygen interaction was fitted to the diffusion Monte Carlo calculations for water molecules with the two-legged configuration adsorbed on the graphene sheet \cite{Brandenburg2019Physisorption}; we put the molecular force on oxygen atoms. The interaction has the Morse potential shape (see Fig. \ref{potential}): 
\begin{equation} \label{g-O} E_b^{\mathrm{O}}(d) = D_e^{\mathrm{O}} \left[ \left( 1 - e^{-a^{\mathrm{O}}(d-d_0^{\mathrm{O}})}  \right)^2 -1 \right], 
\end{equation} 
where $D_e^{\mathrm{O}} = 9.55185$ kJ mol$^{-1}$, $a^{\mathrm{O}} = 1.34725$ {\AA}$^{-1}$, and $d_0^{\mathrm{O}} = 3.37265$ {\AA} \cite{Brandenburg2019Physisorption}. 

We tried several force potentials to fit the graphene-CO$_2$ interaction using the optB86-vdW DFT data reported in \cite{Takeuchi2017Adsorption} (see Fig. \ref{potential}),
and found that the Morse potential best reproduces the interaction, 
\begin{equation} \label{g-CO2}
E_b^{\mathrm{CO_2}}(d) = D_e^{\mathrm{CO_2}} \left[ \left( 1 - e^{-a^{\mathrm{CO_2}}(d-d_0^{\mathrm{CO_2}})}  \right)^2 -1 \right],
\end{equation} 
where $D_e^{\mathrm{CO_2}} = 21.50492$ kJ mol$^{-1}$, $a^{\mathrm{CO_2}} = 1.18449$ {\AA}$^{-1}$, and $d_0^{\mathrm{CO_2}} = 3.25282$ {\AA}. 
The graphene-carbon interaction was then obtained by
\begin{equation} \label{bare_C}
E_b^{\mathrm{C}} = E_b^{\mathrm{CO_2}} - 2E_b^{\mathrm{O}}. 
\end{equation} 
However, this gives a nonphysical attractive potential at short distances ($<$3 \AA) (Fig. \ref{potential}(b)). To correct this, we added the following repulsive term to the graphene-carbon interaction: 
\begin{equation} \label{repulsive}
E_b^{\mathrm{rep}} = D_e^{\mathrm{rep}} e^{-a^{\mathrm{rep}} \left(d - d_0^{\mathrm{rep}} \right)},
\end{equation} 
where $D_e^{\mathrm{rep}} = 18.69565$ kJ mol$^{-1}$, $a^{\mathrm{rep}} = 26.09666$ {\AA}$^{-1}$, and $d_0^{\mathrm{rep}} = 2.56637$ {\AA}.
The overall graphene-carbon interaction is 
\begin{equation} \label{g-C} 
E_b^{\mathrm{C}}(d) = D_e^{\mathrm{CO_2}} \left[ \left( 1 - e^{-a^{\mathrm{CO_2}}(d-d_0^{\mathrm{CO_2}})}  \right)^2 - 1 \right] - 2D_e^{\mathrm{O}} \left[ \left( 1 - e^{-a^{\mathrm{O}}(d-d_0^{\mathrm{O}})}  \right)^2 -1 \right] + D_e^{\mathrm{rep}} e^{-2 a^{\mathrm{rep}} (d-d_0^{\mathrm{rep}})}, 
\end{equation}  
(Fig. \ref{potential}(b)). There is no interaction between graphene and hydrogen atoms in our simulations. We implemented the interaction Eqs. (\ref{g-O},\ref{g-C}) in the Qbox code.

\clearpage
To study graphene-confined solutions, we modeled two model graphene surfaces in the $xy$ plane by defining the $z$-dependent potentials on oxygen and carbon atoms (Eq. (\ref{g-O}) and (\ref{g-C})). Water and CO$_2$ molecules were inserted between these sheets. 
There is 7 {\AA} thick vacuum, so 
the confined solution is at least 10 {\AA} away from its replica with periodic boundary conditions.
The $x$- and $y$-dimensions of the unit cell are the same as those with stishovite confinement.
We calculated the inhomogeneous particle density of solutions along the $z$-direction to get the actual thickness of confined solutions, 
and then we can obtain the actual volume of solutions, which can be used to calculate the solution pressures with the equation of state \cite{Duan2006Equation} (see the main text).

We also validated pressures using density functional theory. 
We calculated the lateral pressure in the simulation cell: $P_{\parallel}= \left( \sigma'_{xx} + \sigma'_{yy} \right) / 2$. The diagonal elements of the computed stress tensor, $\sigma_{xx}$ and $\sigma_{yy}$, are modified to account for the vacuum in the unit cell according to $\sigma_{i}' = \sigma_{i} \left( L_z / h_z \right)$, where $L_z$ is the $z$-dimention of the unit cell, and $h_z$ is the distance between graphene sheets.
This method was used in previous studies \cite{Chen2016Two}.

For stishovite-confined solutions, we inserted a slab of SiO$_2$, which is made by three stoichiometric layers of stishovite exposing the low-energy (100) surface \cite{Feya2018Tetrahedral} in the $xy$ plane, into the cell. 
Table \ref{simulations-table} summarizes the simulation setups.

\vfill
\begin{table}[ht]
\caption{Details of the AIMD simulations. x(CO$_2$) is the initial mole fraction of CO$_2$(aq). $N_{\mathrm{CO_2}}$ and $N_{\mathrm{H_2O}}$ are the initial number of CO$_2$ and H$_2$O molecules, respectively, in the unit cell. $h_z$ is the distance between two confining graphene sheets. $P_{\parallel}$ is the lateral pressure: $ P_{\parallel} = \left( \sigma'_{xx} + \sigma'_{yy} \right) / 2$. 
$ P_{\perp} $ is the pressure along the $z$ direction, $P_{\perp} = \sigma_{zz}$. The uncertainties are standard deviations.}
\label{simulations-table}
\centerline{
\begin{tabular}{ c c c c c c c c }
\hline 
\hline
$\:$Confinement$\:$ & $\:$x(CO$_2$)$\:$ & $\:$ $N_{\mathrm{CO_2}}$ $\:$ & $\:$ $N_{\mathrm{H_2O}}$ $\:$ & $\:$ T (K) $\:$ & $\:$ $h_z$ ({\AA})$\:$ & $\:$ P (GPa) $\:$ & $\:$ $\rho$ (kg m$^{-3}$) $\:$ \\
\hline
Graphene     & 0.185          & 5                     & 22                    & 1000 & 9.0 & $P_{\parallel} = 10.9 \pm 1.4$   & 1707 \\
Graphene     & 0.185          & 5                     & 22                    & 1400 & 9.2 & $P_{\parallel} = 9.4 \pm 1.5$     & 1622 \\
SiO$_2$      & 0.185          & 5                     & 22                    & 1000 & 6.9 & $P_{\perp} = 9.5 \pm 2.0$ &     -          \\
SiO$_2$      & 0.185          & 5                     & 22                    & 1400 & 7.1 & $P_{\perp} = 9.8 \pm 2.2$ &     -          \\
\hline
\hline
\end{tabular}
}
\end{table}

\vfill
\clearpage
\begin{figure}[ht]
\centering
\includegraphics[width=0.7\textwidth]{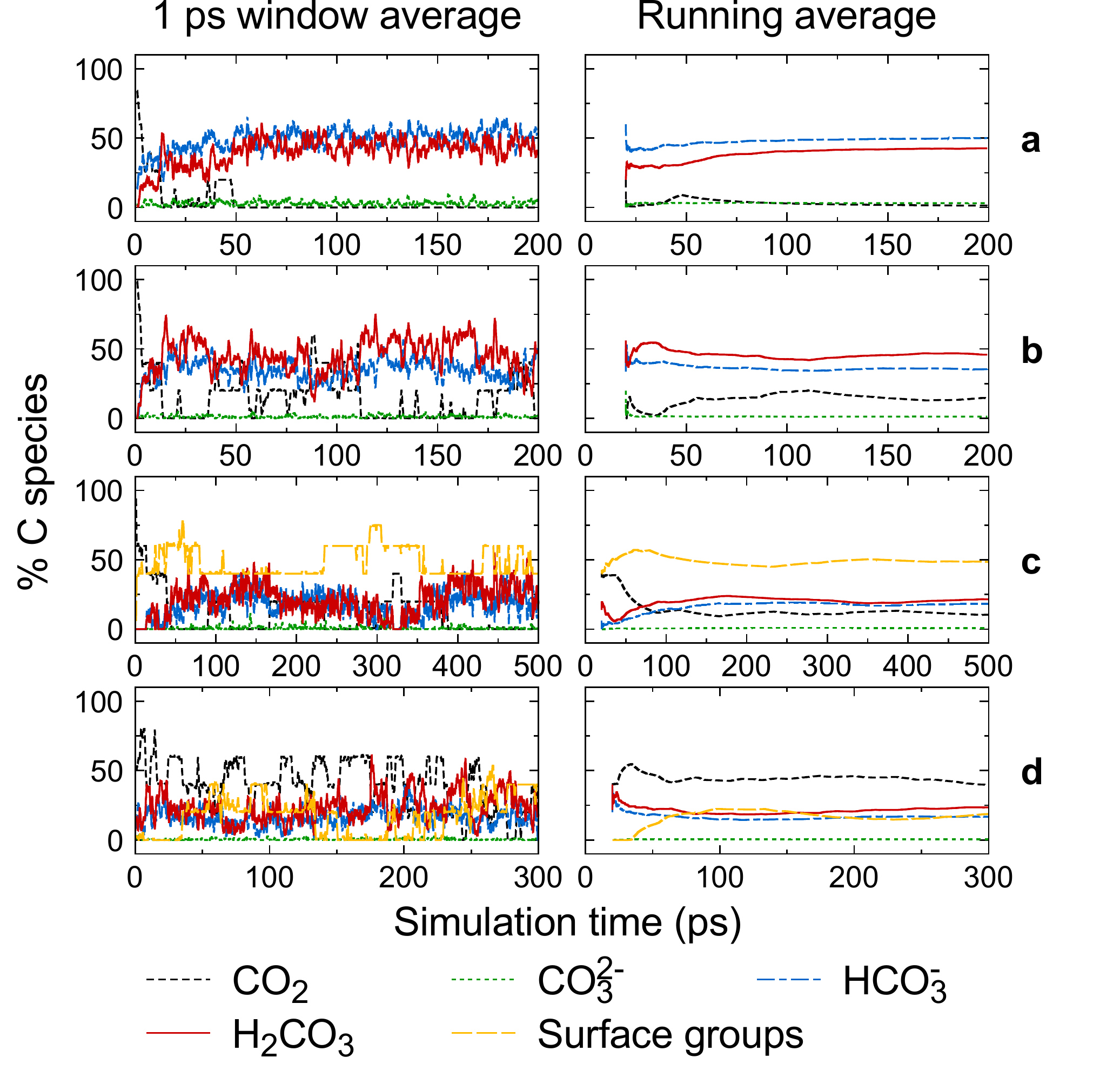}
\caption{ Mole percents of carbon species as functions of simulation time in AIMD simulations. Left panel: the 1 ps window average. Right panel: the running average excluding the first 20 ps equilibration.
(\textbf{a}) x(CO$_2$) = 0.185, 1000 K, graphene confinement. (\textbf{b}) x(CO$_2$) = 0.185, 1400 K, graphene confinement. (\textbf{c}) x(CO$_2$) = 0.185, 1000 K, stishovite confinement. (\textbf{d}) x(CO$_2$) = 0.185, 1400 K, stishovite confinement. }
\label{convergence}
\end{figure}

\clearpage
\begin{figure}[ht]
\centering
\includegraphics[width=0.7\textwidth]{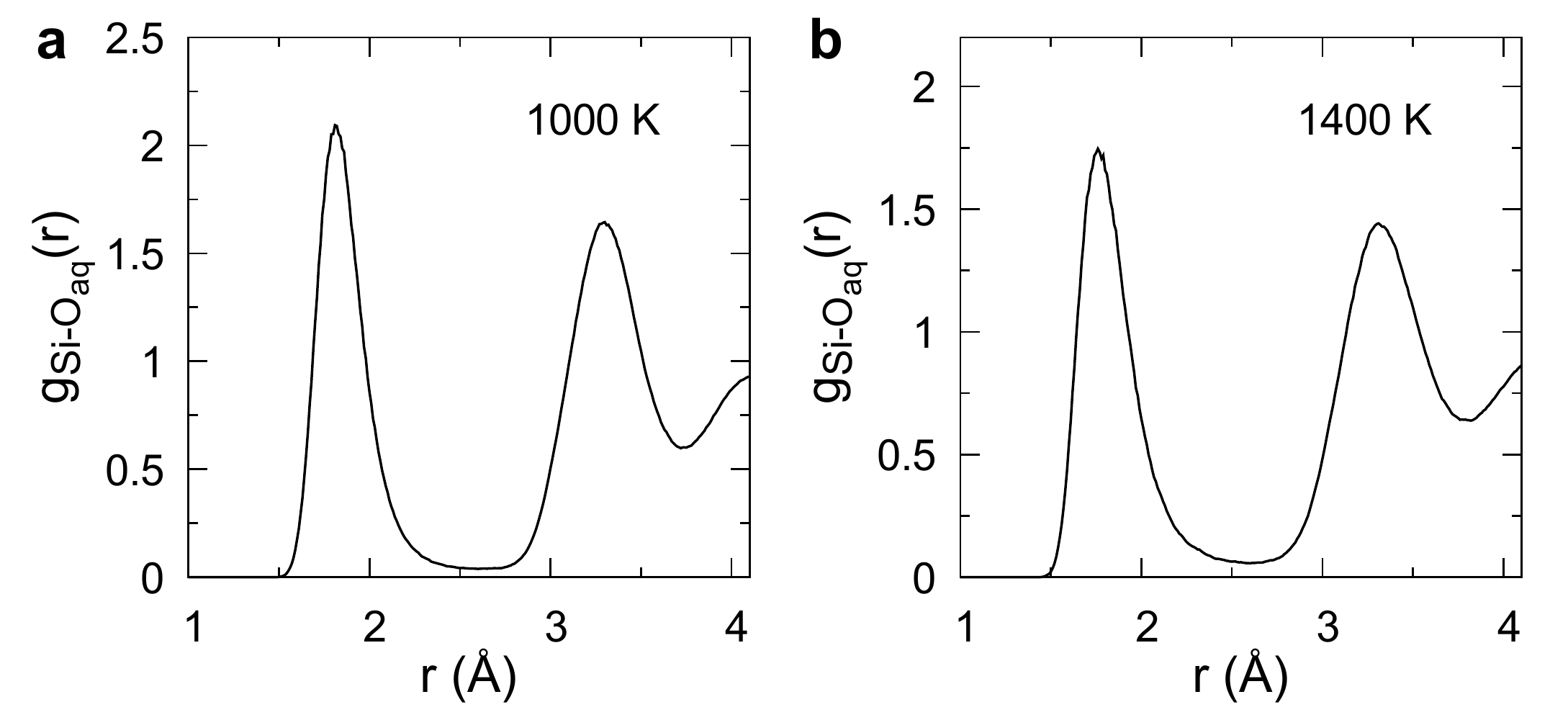}
\caption{Radial distribution functions (RDFs) of the silicon atoms (Si) in the solid-liquid interface layer of stishovite slabs versus the oxygen atoms (O$_{aq}$) in the solutions. 
(\textbf{a}) 1000 K and (\textbf{b}) 1400 K.
The initial mole fraction of  CO$_2$(aq) is 0.185. 
The oxygen atoms in the first peak of RDFs are considered as bonded to the silicon atoms in stishovite.}
\label{si-ow-rdf}
\end{figure}

\clearpage
\begin{figure}[ht]
\centering
\includegraphics[width=0.45\textwidth]{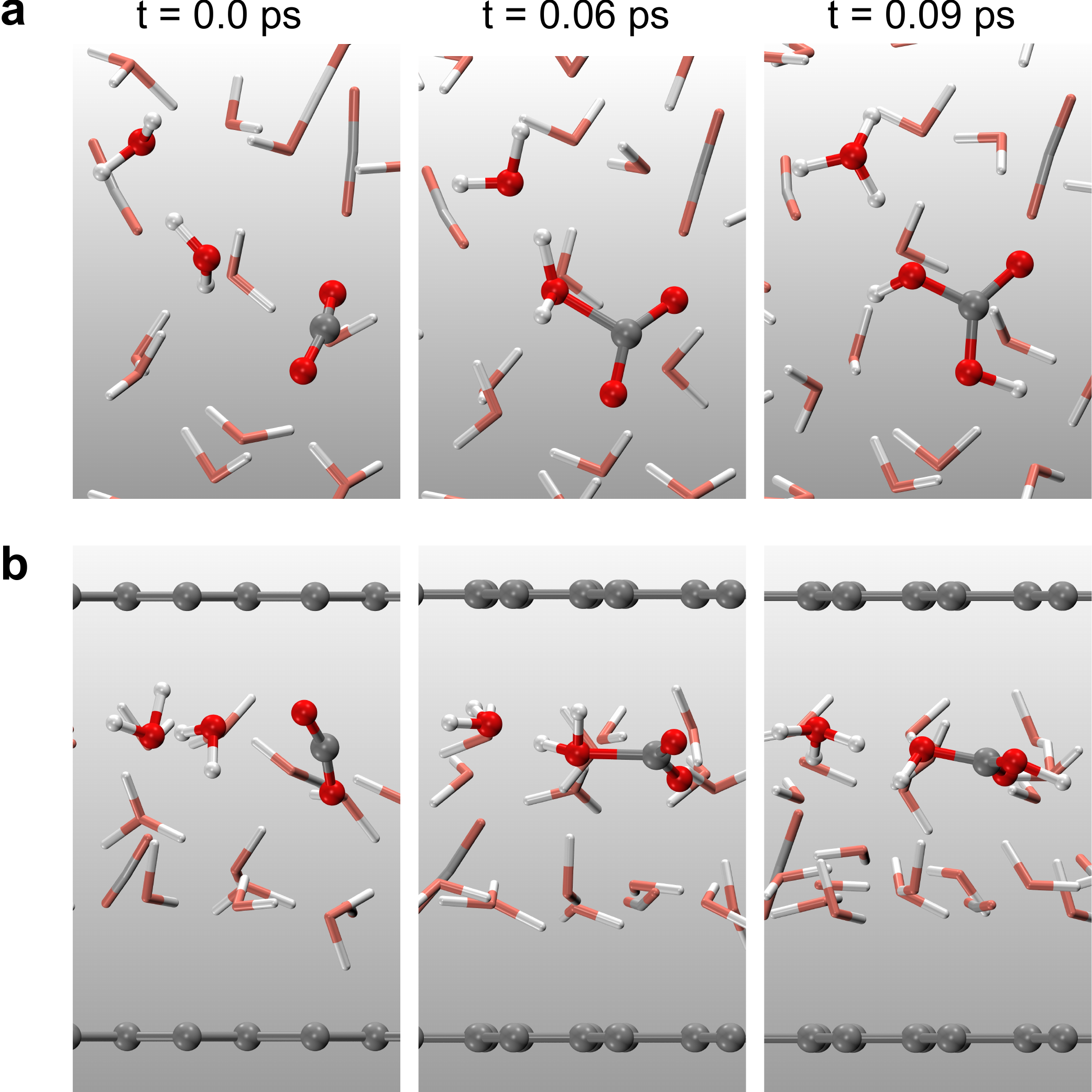}
\caption{Reactions between water and CO$_2$(aq) nanconfined by graphene. (\textbf{a}) Top view; (\textbf{b}) Side view.}
\label{reaction-graphene}
\end{figure}

\begin{figure}[ht]
\centering
\includegraphics[width=0.5\textwidth]{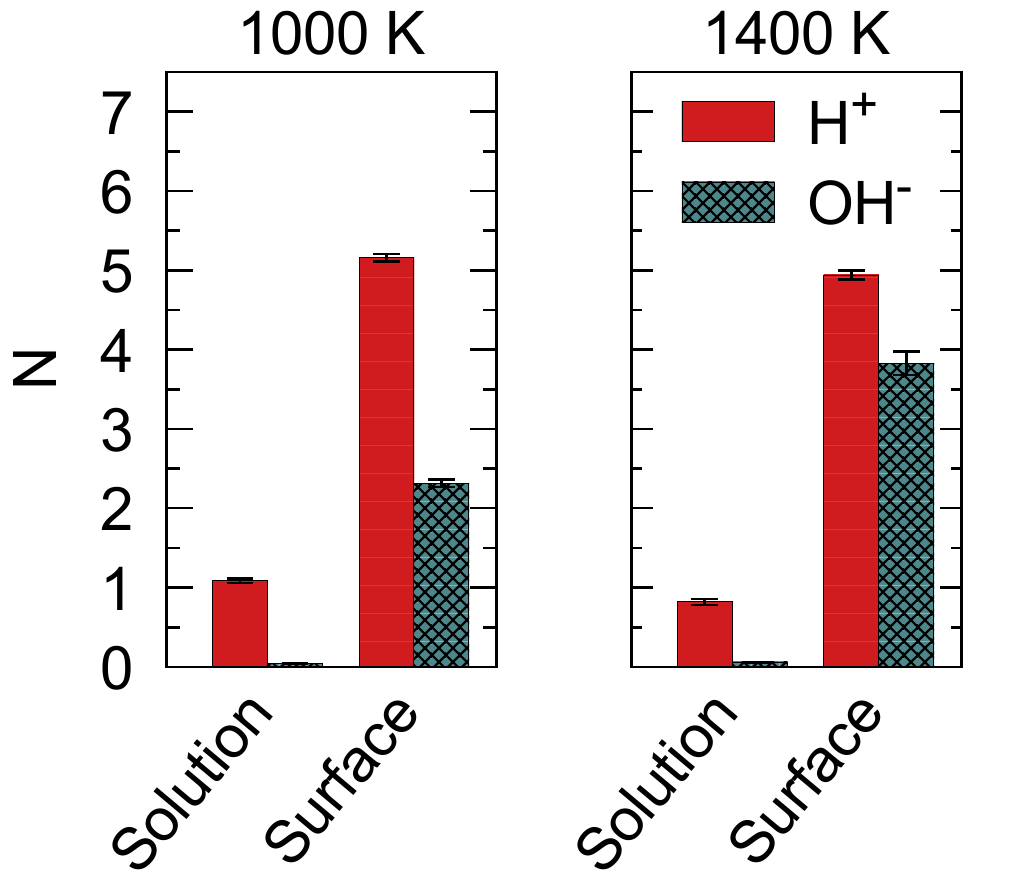}
\caption{ Number ($N$) of H$^+$ and OH$^-$ ions dissolved in the solutions and adsorbed at the stishovite surface in one unit cell.
The initial mole fraction of CO$_2$(aq) is 0.185. The pressure is $\sim$10 GPa, and the temperature is 1000 K (left) and 1400 K (right).}
\label{water-groups}
\end{figure}

\clearpage

\begin{table}[ht]
\caption{Equilibrium percents of carbon species in total dissolved carbon in solutions. The initial mole fraction of CO$_2$(aq) is 0.185. The pressure is $\sim$10 GPa, and the temperature is 1000 K. Bulk values are taken from Ref. \cite{Stolte2019Large}.}
\label{species2}
\centerline{
\begin{tabular}{ c  c  c  c }
\hline 
\hline
$\quad$Confinement$\quad$ & $\quad$Bulk$\quad$ & $\quad$Graphene$\quad$ & $\quad$ Stishovite $\quad$ \\
\hline
CO$_2$           & 15.1 $\pm$ 2.0 & 1.3 $\pm$ 0.9  & 10.5 $\pm$ 2.3  \\
CO$_3^{2-}$      & 1.1 $\pm$ 0.0  & 3.1 $\pm$ 0.1  & 0.7 $\pm$ 0.1   \\
HCO$_3^-$        & 35.9 $\pm$ 0.7 & 50.0 $\pm$ 1.0 & 18.2 $\pm$ 1.8  \\
H$_2$CO$_3$      & 46.8 $\pm$ 1.5 & 42.7 $\pm$ 1.7 & 21.6 $\pm$ 2.7  \\
H$_3$CO$_3^+$    & 0.8 $\pm$ 0.1  & 1.3 $\pm$ 0.1  & 0.5 $\pm$ 0.1   \\
Pyrocarbonate    & 0.3 $\pm$ 0.1  & 1.6 $\pm$ 0.8  & 0.0             \\
Surface groups   & -              & -              & 48.5 $\pm$ 2.1  \\
\hline
\hline
\end{tabular}
}
\end{table}

\begin{table}[ht]
\caption{Equilibrium percents of carbon species in total dissolved carbon in solutions. The initial mole fraction of CO$_2$(aq) is 0.185. The pressure is $\sim$10 GPa, and the temperature is 1400 K. The bulk values are obtained by interpolating the results 
for x(CO$_2$) = 0.032, 0.143 and 0.333 from Ref. \cite{Stolte2019Large} with cubic splines.}
\label{species3}
\centerline{
\begin{tabular}{ c | c  c  c }
\hline 
\hline
$\quad$Confinement$\quad$ & $\quad$Bulk $\quad $ & $\quad$Graphene$\quad$ & $\quad$ SiO$_2$ $\quad$ \\
\hline
CO$_2$           & 58.8 $\pm$ 2.0  & 14.5 $\pm$ 3.2 & 39.8 $\pm$ 3.6  \\
CO$_3^{2-}$      & 0.1 $\pm$ 0.0   & 1.3 $\pm$ 0.1  & 0.5 $\pm$ 0.1   \\
HCO$_3^-$        & 16.8 $\pm$ 0.7  & 35.5 $\pm$ 1.0 & 16.6 $\pm$ 1.2  \\
H$_2$CO$_3$      & 23.6 $\pm$ 1.4  & 45.9 $\pm$ 2.3 & 23.3 $\pm$ 1.7  \\
H$_3$CO$_3^+$    & 0.3 $\pm$ 0.1   & 2.0 $\pm$ 0.6  & 0.6 $\pm$ 0.0   \\
Pyrocarbonate    & 0.4 $\pm$ 0.1   & 0.7 $\pm$ 0.3  & 0.2 $\pm$ 0.1   \\
Surface groups   & -               & -              & 19.1 $\pm$ 3.1  \\
\hline
\hline
\end{tabular}
}
\end{table}

\clearpage
\bibliography{ref}